\documentclass[twocolumn,prl,superscriptaddress,floatfix,aps]{revtex4-1}

\usepackage{amsmath}
\usepackage{amssymb}
\usepackage{tabularx}
\usepackage{makecell}
\usepackage[dvipsnames]{xcolor}
\usepackage{graphicx}
\usepackage{bm}
\usepackage{siunitx}
\usepackage{hyperref}
\usepackage{chemformula}

\usepackage{xcolor}
\hypersetup{
    colorlinks,
    linkcolor={red!50!black},
    citecolor={blue!50!black},
    urlcolor={blue!80!black}
}

\newcommand{\BTI}{BiTeI}
\newcommand{\BBTI}{Bi$_2$TeI}
\newcommand{\BBBTI}{Bi$_3$TeI}

\begin{document}
\title{Pressure-induced superconductivity in the weak topological insulator Bi$_2$TeI and the topological metal Bi$_3$TeI}
\author{T.\ A.\ Elmslie}
\affiliation{Department of Physics, University of Florida, Gainesville, FL 32611, USA}
\author{D.\ VanGennep}
\affiliation{Department of Physics, University of Florida, Gainesville, FL 32611, USA}
\author{R.\ N.\ Baten}
\affiliation{Department of Physics, University of Florida, Gainesville, FL 32611, USA}
\author{J.\ Downing}
\affiliation{Department of Physics, University of Florida, Gainesville, FL 32611, USA}
\author{W.\ Bi}
\affiliation{Advanced Photon Source, Argonne National Laboratory, Argonne, IL 60439, USA}
\affiliation{Department of Geology, University of Illinois at Urbana-Champaign, Urbana, IL 61801, USA}
\affiliation{Department of Physics, University of Alabama at Birmingham, Birmingham, AL, 35294, USA}
\author{S.~T.~Weir}
\affiliation{Physics Division, Lawrence Livermore National Laboratory, Livermore, CA 94550, USA}
\author{Y.\ K.\ Vohra}
\affiliation{Department of Physics, University of Alabama at Birmingham, Birmingham, AL, 35294, USA}
\author{R.\ E.\ Baumbach}
\affiliation{National High Magnetic Field Laboratory, Florida State University, Tallahassee, Florida 32310, USA}
\affiliation{Department of Physics, Florida State University, Tallahassee, Florida 32306, USA}
\author{J.\ J.\ Hamlin}
\email{Corresponding author: jhamlin@ufl.edu}
\affiliation{Department of Physics, University of Florida, Gainesville, FL 32611, USA}

\begin{abstract}
We report a series of high-pressure electrical transport, magnetic susceptibility, and x-ray diffraction measurements on single crystals of the weak topological insulator Bi$_2$TeI and the topological metal Bi$_3$TeI.
Room temperature x-ray diffraction measurements show that both materials go through a series of pressure-induced structural transitions and eventually adopt a disordered $bcc$ alloy structure at high pressure.
A re-analysis of the published data on BiTeI indicates that this material also adopts a disordered $bcc$ structure at high pressure, in contrast to the previously suggested $P4/nmm$ structure.
We find that Bi$_2$TeI and Bi$_3$TeI become superconducting at $\sim$\SI{13}{\giga \pascal} and $\sim$\SI{11.5}{\giga \pascal}, respectively.
The superconducting critical temperature $T_c$ of the $bcc$ phase reaches maximum values of \SI{7}{\kelvin} and \SI{7.5}{\kelvin} for Bi$_2$TeI and Bi$_3$TeI, respectively and $dT_c/dP < 0$ in both cases.
The results indicate that disordered alloy $bcc$ superconducting phases appear to be a universal feature of both the Bi-Te and Bi-Te-I systems at high pressure.
\end{abstract}

\maketitle

\section{Introduction}
\vspace{-0.5em}
Several bismuth-based compounds have gained attention due to their diverse topologically nontrivial electronic properties~\cite{hasan_colloquium_2010,ando_topological_2013,qi_topological_2011,wehling_dirac_2014}. 
Interest in the BiTe$X$ ($X$ = Cl, Br, I) family of compounds first surged when it was discovered, via spin- and angle-resolved photoemission spectroscopy, that BiTeI displays an enormous Rashba-like spin splitting of the bulk electronic bands~\cite{ishizaka_giant_2011}.
This result was soon followed by similar observations in BiTeBr and BiTeCl~\cite{sakano_strongly_2013,landolt_bulk_2013}. 

Compounds in the BiTe$X$ family have been shown to exhibit drastic changes in their physical properties when external pressure is applied. 
BiTeI and BiTeBr are believed to undergo pressure-driven topological quantum phase transitions~\cite{bahramy_emergence_2012,rusinov_pressure_2016,vangennep_pressure_2014,xi_signatures_2013,park_quantum_2015,crassee_bitecl_2017}, while all of the compounds have shown evidence of pressure-induced structural transitions~\cite{crassee_bitecl_2017,chen_high-pressure_2013,sans_structural_2016} and pressure-induced superconductivity~\cite{vangennep_pressure-induced_2017,qi_topological_2017,jin_superconductivity_2017,ying_realization_2016,jin_pressure-induced_2017}.
Compounds with the formula \BBTI{}~\cite{savilov_new_2005,babanly_phase_2009} and \BBBTI{}~\cite{zeugner_modular_2017} have also been synthesized.
These compounds incorporate Bi-bilayers in the van der Waals gap between layers of BiTeI~\cite{zeugner_modular_2017,ryu_growth_2016,tu_thermoelectric_2018,aliev_solid-state_2008,babanly_phase_2009,savilov_new_2005,rusinov_mirror-symmetry_2016}.
Single crystals of Bi$_2$TeBr and Bi$_3$TeBr have also been synthesized and have been shown to be isostructural with \BBTI{} and \BBBTI{}, respectively~\cite{zeugner_synthesis_2018}. 

According to parity eigenvalues, bulk \BBTI{} has been shown to belong to the $\mathbb{Z}_2$ (0,001) class of 3D weak topological insulators (TIs) under ambient conditions, and consists of alternating stacks of quantum spin Hall layers and normal insulator (NI) layers~\cite{tang_weak_2014}.
This structure has been predicted to give rise to two fairly isotropic Dirac cones that exist on the side surfaces, perpendicular to the BiTeI planes.
These 3D weak TIs have recently gained interest due to their potential application as high-performance thermoelectric materials, partially due to the possibility of achieving minimum lattice thermal conductivity.
This minimum lattice thermal conductivity occurs when phonons have mean free paths on the order of one phonon wavelength. 
This has been experimentally observed in the 3D weak TI Bi$_{14}$Rh$_3$I$_9$ and has been theoretically predicted in \BBTI{}~\cite{wei_minimum_2016}.
\BBBTI{} may also exhibit unusual topological properties. 
Previous work~\cite{zeugner_modular_2017} found evidence of band inversion when spin-orbit coupling is accounted for as well as an unconventional surface state that resides on various termination layers. 
However, the inverted gap does not occur at a time reversal invariant momentum point and should not create a topological surface state according to the Fu-Kane $\mathbb{Z}_2$ classification~\cite{kane_z_2005}. 

Many topological materials have been reported to exhibit pressure-induced superconductivity, though the superconducting phase typically emerges following a structural phase transition that modifies the topological properties~\cite{kirshenbaum_pressure-induced_2013,he_pressure-induced_2016,kang_superconductivity_2015,qi_superconductivity_2016}.
In this work, we report a series of electrical resistivity, magnetic susceptibility, and x-ray diffraction measurements on \BBTI{} and \BBBTI{} to pressures as high as $\sim$\SI{30}{\giga \pascal}. 
We find that both compounds go through a series of pressure-induced structural transitions, with the high pressure phases exhibiting a disordered $bcc$ structure very similar to that previously observed for other Bi-Te compounds at high pressure~\cite{stillwell_superconducting_2016}.
In both compounds, the high pressure $bcc$ phase is superconducting, with a maximum $T_c$ value near \SI{7}{K}.
We observe near 100\% flux expulsion in AC magnetic susceptibility measurements, consistent with bulk superconductivity.

\section{Methods}
The synthesis of both \BBTI{} and \BBBTI{} were guided by the results found in Ref.~\cite{zeugner_modular_2017}.
Polycrystalline \BBTI{} was initially grown by solid state reaction of stoichiometric amounts of the individual elements sealed under vacuum in a quartz tube, and annealed at \SI{425}{\celsius} for 6 days. 
Ambient pressure x-ray diffraction measurements of the polycrystalline material showed phase-pure \BBTI. The polycrystalline \BBTI{} was then ground into a fine powder, sealed under vacuum in a quartz tube, and subjected to a horizontal temperature gradient of $390-\SI{410}{\celsius}$ for a period of 19 days.
Single crystals of \BBTI{} nucleated on the source material in the hot zone.
These crystals were hexagonal platelets, with typical dimensions of $200-\SI{500}{\micro \meter}$.
The structure of the single crystals was confirmed via powder x-ray diffraction.
Rietveld refinement with the space group $C2/m$ (\#12) yielded lattice constants of a = 7.58 \AA, b = 4.38 \AA, c = 17.74 \AA, and $\beta$ = 98.20$^{\circ}$, which are  consistent with values reported in the literature~\cite{aliev_solid-state_2008,babanly_phase_2009,zeugner_modular_2017,rusinov_mirror-symmetry_2016}, although there are slight discrepancies in the literature about which space group should be assigned to this compound. 
Reference~\cite{rusinov_mirror-symmetry_2016} noted that applying a crystallographic transformation matrix to the unit cell yields a reduced trigonal rhombohedral cell with a = b = 4.38 \AA, c = 17.74 \AA, $\alpha$ = $\beta$ = 82.905$^{\circ}$, and $\gamma$ = 60.002$^{\circ}$. 

The \BBBTI{} samples were grown in a Bi self-flux, following the method described in Ref.~\cite{zeugner_modular_2017}.
This resulted in cm-sized single crystals of \BBBTI{}, and the structure was confirmed via both powder and single crystal x-ray diffraction.
Rietveld refinement confirmed the polar noncentrosymmetric space group $R3m$ (\#160) with lattice constants of a = b = 4.40 \AA{} and c = 32.23 \AA, which are  consistent with those reported in the literature~\cite{zeugner_modular_2017}.

Angle-dispersive x-ray diffraction (XRD) experiments on powdered samples were carried out at beamline 16-ID-B, Advanced Photon Source (APS), Argonne National Laboratory.
The x-ray beam had dimensions of approximately \SI{15}{\micro \meter} $\times$ \SI{15}{\micro \meter} and wavelength \SI{0.4066}{\angstrom}.
Powdered CeO$_2$ was used to calibrate the distance and tilting of the detector.
For the x-ray diffraction measurements, high pressure was achieved in Mao-type symmetric DACs (diamond anvils cells) with c-BN seats to allow access to high diffraction angles.
A mortar and pestle were used to powder each sample, which was then loaded into a Symmetric DAC alongside a piece of platinum foil and a ruby flake.
During measurement, platinum and ruby were used to determine pressure within the cell.
The cell containing the Bi$_2$TeI sample used diamonds with a culet diameter of \SI{0.6}{\milli \meter} and a gasket made from SS 316 featuring a hole of approximate diameter \SI{250}{\micro \meter}.
The cell used for measuring Bi$_3$TeI contained diamonds with a culet diameter of \SI{0.5}{\milli \meter}.  Its gasket was also made of SS 316, and had a \SI{200}{\micro \meter} diameter hole.
In both cells, the gasket hole was filled with a pressure medium of 1:1 n-pentane isoamyl alcohol, and pressure was applied in-situ using a computer-controlled gearbox.
The resulting diffraction patterns were processed into usable XRD patterns with Dioptas software~\cite{prescher_dioptas_2015} and analyzed with GSAS-II software~\cite{toby_gsas-ii_2013} to obtain lattice constants and structure.

Ambient pressure transport properties were measured in a Quantum Design Physical Property Measurement System (PPMS) with temperatures ranging from $2 - \SI{300}{K}$ and magnetic fields up to \SI{9}{T}.
Additional measurements on \ch{Bi3TeI} were performed down to \SI{0.5}{K}.
For the resistivity measurements under pressure, single crystals with typical dimensions of $\sim \SI{70}{\micro m} \times \SI{70}{\micro m} \times \SI{10}{\micro m}$ were cut from larger crystals and loaded into an OmniDAC membrane-driven diamond anvil cell from Almax-EasyLab.
This cell was placed inside of a custom continuous flow cryostat built by Oxford Instruments, which has an optical window at the bottom that allows for \textit{in-situ} pressure measurement of the R$_1$ fluorescence line of ruby~\cite{chijioke_ruby_2005}.
One of the diamonds used was a designer-diamond anvil containing eight symmetrically arranged tungsten microprobes which are encapsulated in high-purity homoepitaxial diamond~\cite{weir_epitaxial_2000}. 
The opposing anvil had a culet size of \SI{500}{\micro \meter}.
The gasket was made of 316 SS, and was preindented to an initial thickness of $\sim \SI{30}{\micro \meter}$. 
Quasihydrostatic, soft, solid steatite was used as the pressure-transmitting medium.
Resistance was measured in the Van der Pauw geometry with a current of $\sim \SI{1}{mA}$.

For the ac susceptibility measurements, one sample of each material was measured in order to test whether the observed superconducting transitions are bulk in nature.
The \BBTI{} sample had dimensions of $\sim \SI{220}{\micro m} \times \SI{200}{\micro m} \times \SI{10}{\micro m}$ and the \BBBTI{} sample had dimensions of $\sim \SI{130}{\micro m} \times \SI{130}{\micro m} \times \SI{30}{\micro m}$. 
The diamond anvil cell used for these measurements was an Almax-EasyLab ChicagoDAC, which is a membrane-driven diamond anvil cell that is made to fit inside the bore of a Quantum Design PPMS~\cite{feng_compact_2014}. 
The ac magnetic susceptibility measurements were performed using a balanced primary/secondary coil system that has been described elsewhere~\cite{deemyad_dependence_2001}. 
Both of the diamonds had 800 $\mu$m culets. 
The gasket was made of Berylco25 which was preindented to an initial thickness of $\sim$60 $\mu$m. 
Samples were loaded into a gasket chamber of $\sim$300 $\mu$m diameter along with small chips of ruby, and the pressure medium used was 1:1 n-pentane:isoamyl alcohol~\cite{torikachvili_solidification_2015}. 
An SR830 lock-in amplifier was used to measure the first harmonic of the ac magnetic susceptibility. 
The primary coil provides an excitation field of 3 Oe rms at $\sim$1 kHz.
The detection coil is connected through a Stanford Research SR554 transformer/preamplifier.
The $R_1$ fluorescence line from the ruby was again used for pressure determination, and was collected via optical fiber and a lens system which is mounted to the diamond anvil cell inside of the cryostat.

For samples labeled S1, pressure was applied at room temperature, then released while keeping the samples below $\sim$\SI{15}{\kelvin}. 
The samples that are labeled S2 were compressed at either 1.8 or \SI{4.5}{\kelvin}, and then decompressed while keeping the samples below $\sim$\SI{15}{\kelvin}.
Samples labeled S3 were slowly compressed and decompressed at room temperature.
For the ac magnetic susceptibility measurements, pressure changes were carried out at room temperature, with the data sets measured during decompression. 

\section{Ambient pressure transport}
\begin{figure}
    \includegraphics[width=0.95\columnwidth]{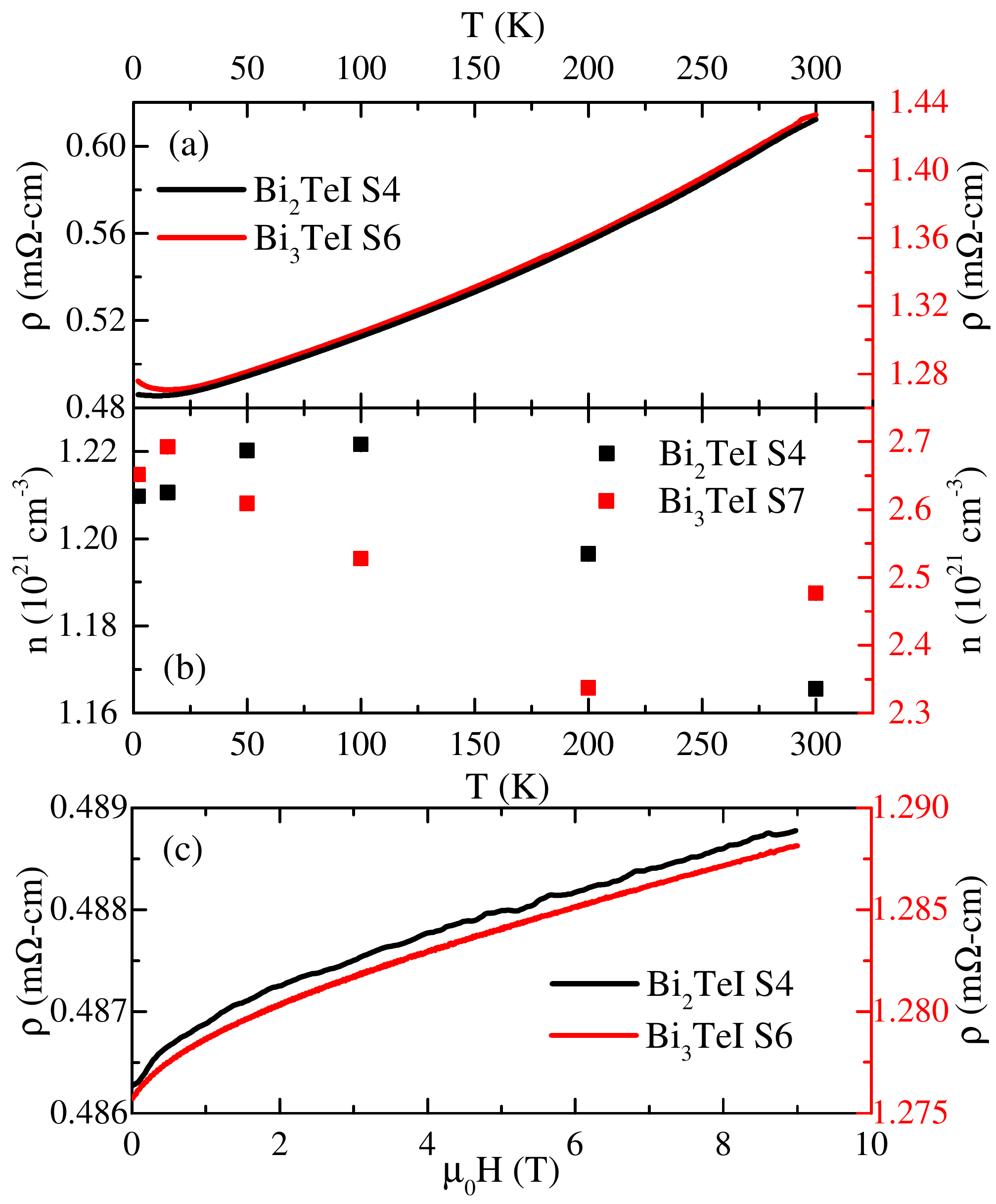}
    \caption{Ambient pressure transport properties measured in the crystallographic ab-plane. The magnetic field was applied along the c-axis. (a) Resistivity vs temperature, (b) Hall carrier density vs temperature, and (c) resistivity vs magnetic field, measured at \SI{2}{\kelvin}.}
 \label{fig:ambP}
\end{figure}

As shown in Fig.~\ref{fig:ambP}, the resistivities of the samples under ambient conditions were on the order of 0.5 $\Omega$-cm for \BBTI{} and 1 m$\Omega$-cm for \BBBTI{}, each of which were weakly temperature dependent. 
In both samples, $\rho$ decreases upon cooling, followed by a slight upturn at low temperature.
Hall data collected at ambient pressure showed that both samples were $n$-type with carrier concentrations of $\sim$1.2x10$^{21}$ cm$^{-3}$ for \BBTI{} and $\sim$2.5x10$^{21}$ cm$^{-3}$ for \BBBTI{}.
The carrier concentrations were also weakly temperature dependent, and showed an increase upon cooling. 
Hall data for both samples up to \SI{9}{\tesla} showed no evidence of third order terms that would indicate multiple types of carriers. 
Magnetoresistivity measurements showed cusps at zero field which may come from weak anitlocalization effects, similar to those observed in the topologically nontrivial semimetal LuPdBi~\cite{xu_weak_2015}.
We see no evidence of superconductivity at ambient pressure above \SI{2}{K} in \ch{Bi2TeI} and above \SI{0.5}{K} in \ch{Bi3TeI}.

\section{resistivity under pressure}
\begin{figure}
    \includegraphics[width=0.95\columnwidth]{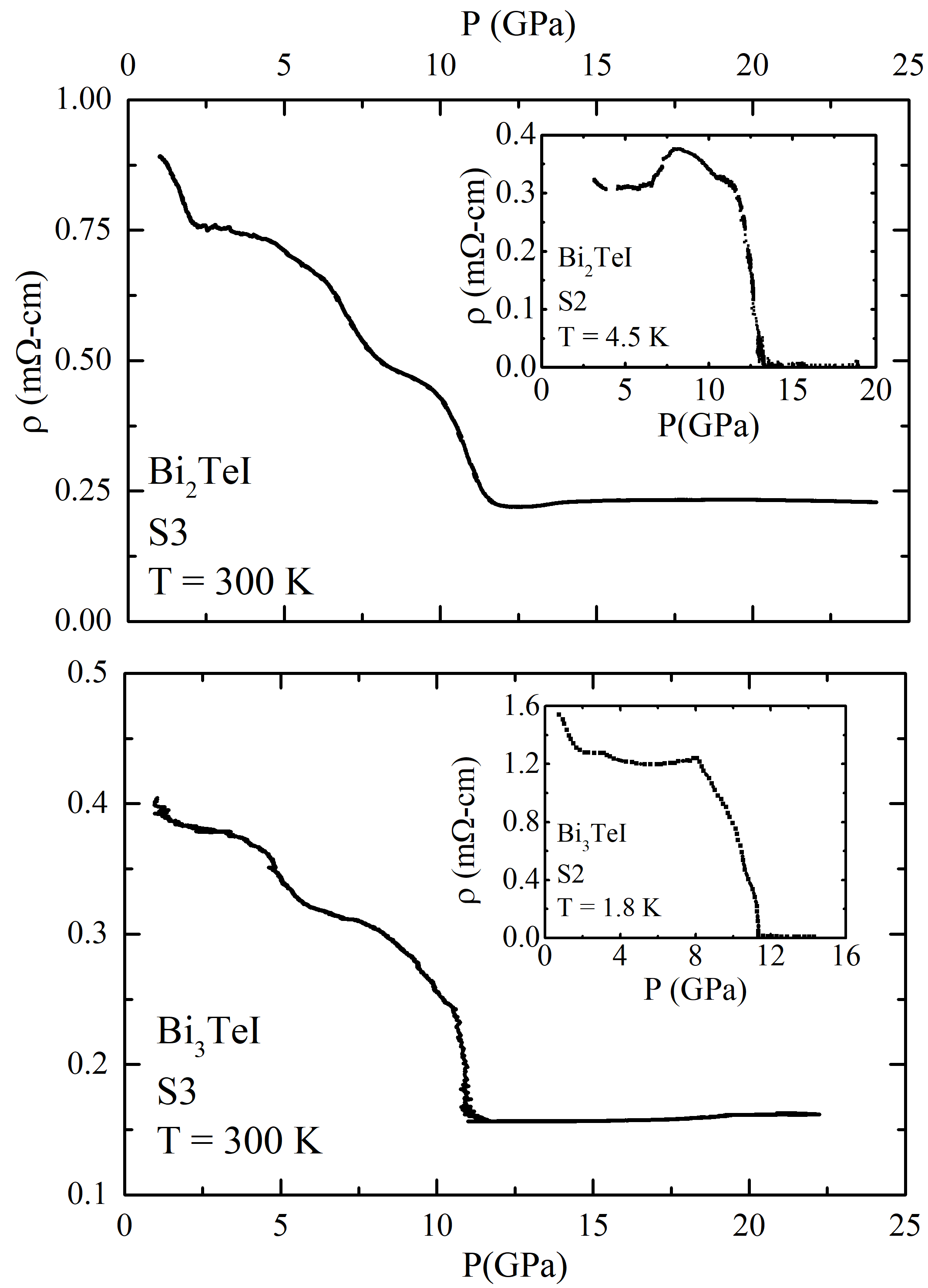}
    \caption{Resistivity as a function of applied pressure at room temperature for sample S3 of \BBTI{} (top). Sample S2 was compressed at \SI{4.5}{\kelvin}, and became superconducting at $\sim$\SI{13}{\giga \pascal} (inset). (bottom) Resistivity as a function of applied pressure at room temperature for samples S3 of \BBBTI{}. Sample S2 was compressed at \SI{1.8}{\kelvin}, and became superconducting at $\sim$\SI{12}{\giga \pascal} (inset).}
 \label{fig:rhovsP}
\end{figure}

Figure~\ref{fig:rhovsP} show how the resistivity of \BBTI{} and \BBBTI{} evolves as a function of pressure.
Samples were compressed isothermally both at room temperature and low temperature (insets).
The room temperature resistivity of both samples show multiple sharp drops as a function of pressure.
The high pressure x-ray scattering results show that these changes in resistivity occur due to pressure-driven structural transitions, similar to what has been observed in \BTI{}~\cite{chen_high-pressure_2013,vangennep_pressure-induced_2017,jin_superconductivity_2017}.
We find significant changes in the resistivity of \BBTI{} occurring between $\sim$\SIrange{6}{8}{\giga \pascal} and $\sim$\SIrange{10}{13}{\giga \pascal}.
In \BBBTI{}, we find similar changes from $\sim$\SIrange{3}{5}{\giga \pascal} and $\sim$\SIrange{8}{11.5}{\giga \pascal}. 
The low temperature isothermal pressure sweeps shown in the insets demonstrate that \BBTI{} begins to superconduct above $\sim$\SI{13}{\giga \pascal}, while \BBBTI{} begins to superconduct above $\sim$\SI{11.5}{\giga \pascal}.
For both materials, we find that the onset of superconductivity appears to coincide with the second (higher pressure) set of features in the room temperature resistivity.
The resistivity of both samples is roughly pressure-independent at pressures higher than the critical pressures needed to induce superconductivity. 

\begin{figure}
    \includegraphics[width=0.8\columnwidth]{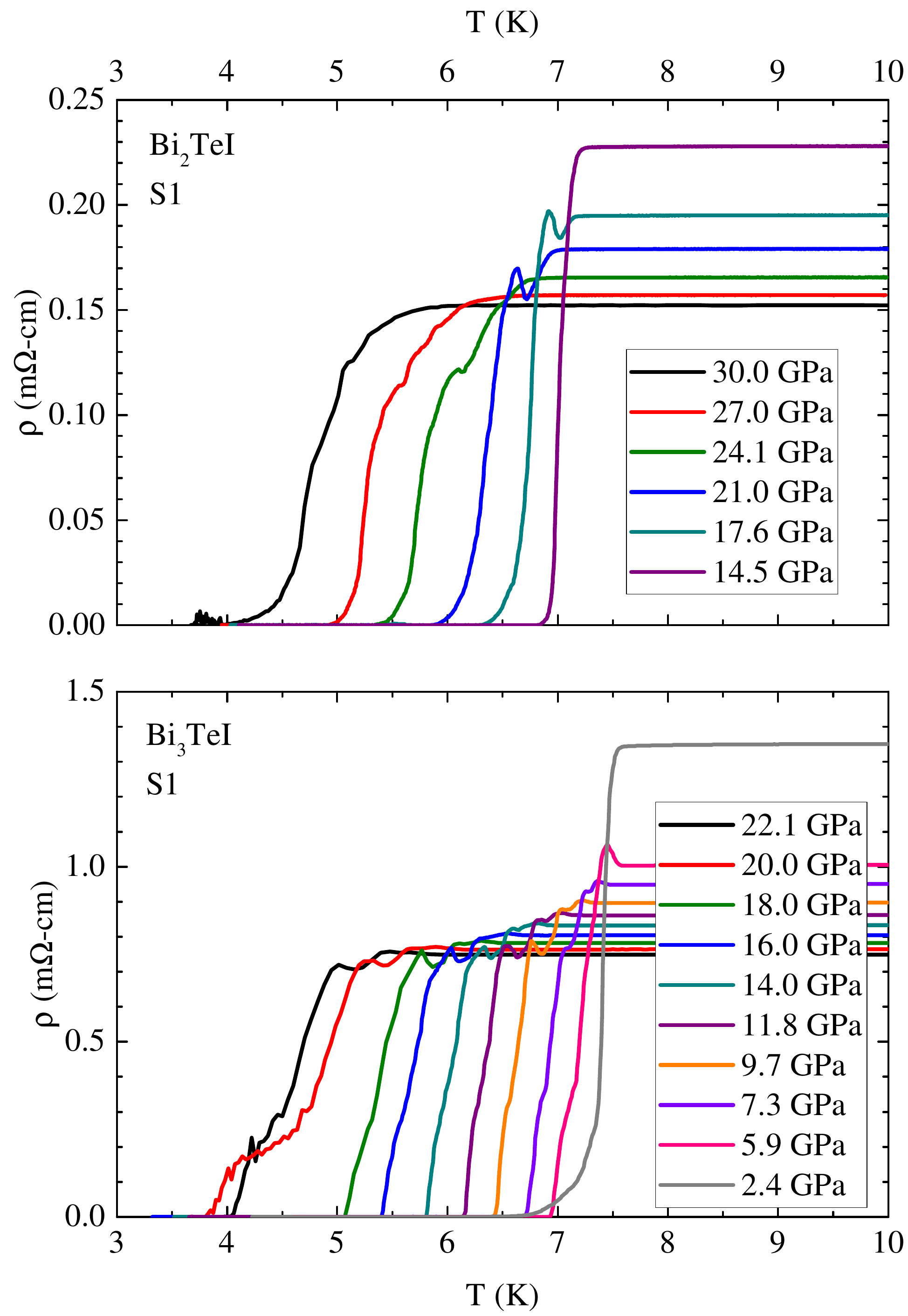}
    \caption{Resistivity as a function of temperature at several different pressures for \BBTI{} (top) and \BBBTI{} (bottom). The data shown were collected upon releasing pressure at low temperature.}
 \label{fig:rhovsT}
\end{figure}

Figure \ref{fig:rhovsT} shows the temperature dependence of the resistivity for \BBTI{} and \BBBTI{} at various pressures.
These data sets were taken upon decompression at low temperature, in which the sample temperature was kept below $\sim$\SI{15}{\kelvin}.
In both samples, we notice an increase in both the magnitude of the normal-state resistivity as well as an increase in T$_c$ upon decompression.
These features are also observed in \BTI{}~\cite{vangennep_pressure-induced_2017}.

\section{magnetic susceptibility}
\begin{figure}
    \includegraphics[width=\columnwidth]{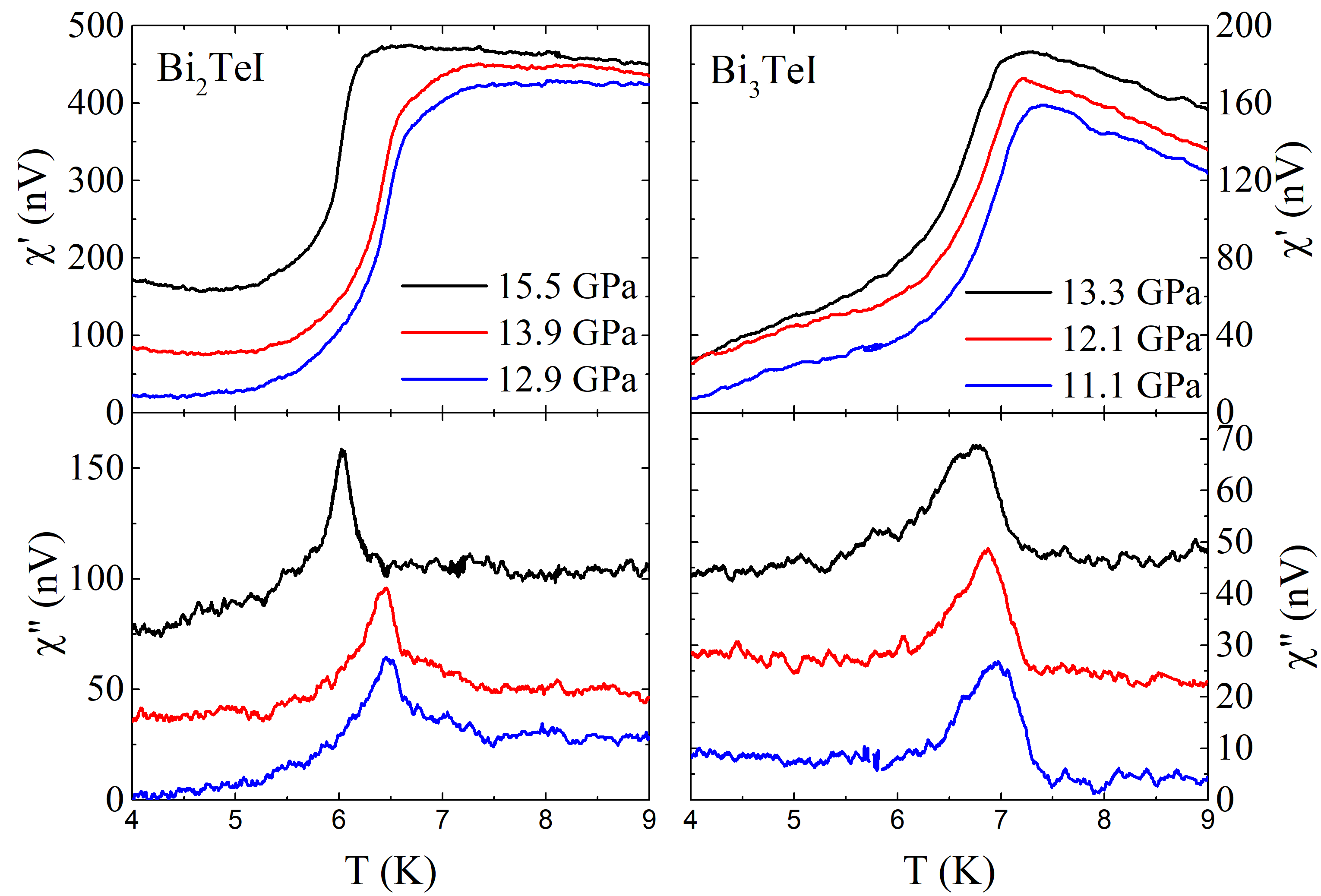}
    \caption{Real ($\chi^\prime$) and imaginary ($\chi^{\prime\prime}$) parts of the AC magnetic susceptibility versus temperature for \BBTI{} at high pressure(left) and \BBBTI{} (right). The observed drops in $\chi^\prime$ are consistent with what is expected from 100\% flux expulsion given the size of the samples, which was calculated to be $\sim$300 nV for \BBTI{} and $\sim$120 nV for \BBBTI{}. The data shown were taken upon releasing pressure.}
 \label{fig:acs}
\end{figure}

It is not uncommon to observe zero resistance in a sample due to small superconducting filaments that exists between voltage or current leads.
Given that elemental Bi, Te, I~\cite{buzea_assembling_2004,hamlin_superconductivity_2015,shimizu_superconductivity_2015}, and a number of Bi-Te~\cite{stillwell_superconducting_2016,jeffries_distinct_2011} phases all become superconducting at high pressure we performed ac magnetic susceptibility measurements in order to determine whether or not the observed superconductivity is intrinsic or due to a small fraction of a minority phase.
Figure \ref{fig:acs} represents a summary of our ac susceptibility experiments.
Upper panels show the temperature dependence of the real part of the susceptibility, $\chi$', while the lower panels show peaks in the imaginary part of the susceptibility, $\chi$''. 
For both materials, the size of the drop in $\chi$' agrees well with the expected value for full magnetic shielding~\cite{nikolo_superconductivity_1995}.
The negative values of $dT_c/dP$ measured via ac susceptibility are consistent with those measured via resistivity with similar slopes. 
For \BBTI{}, we observe that the resistance of sample S2 drops to zero around \SI{4}{\kelvin} at \SI{6}{\giga \pascal}, while we observe no hints of bulk superconductivity via ac susceptibility down to \SI{2}{\kelvin} at roughly the same pressure.
This leads us to conclude that the portions of the sample can persist in the high-pressure phases due to metastability or slow kinetics when pressure is released.

\begin{figure}
    \includegraphics[width=0.8\columnwidth]{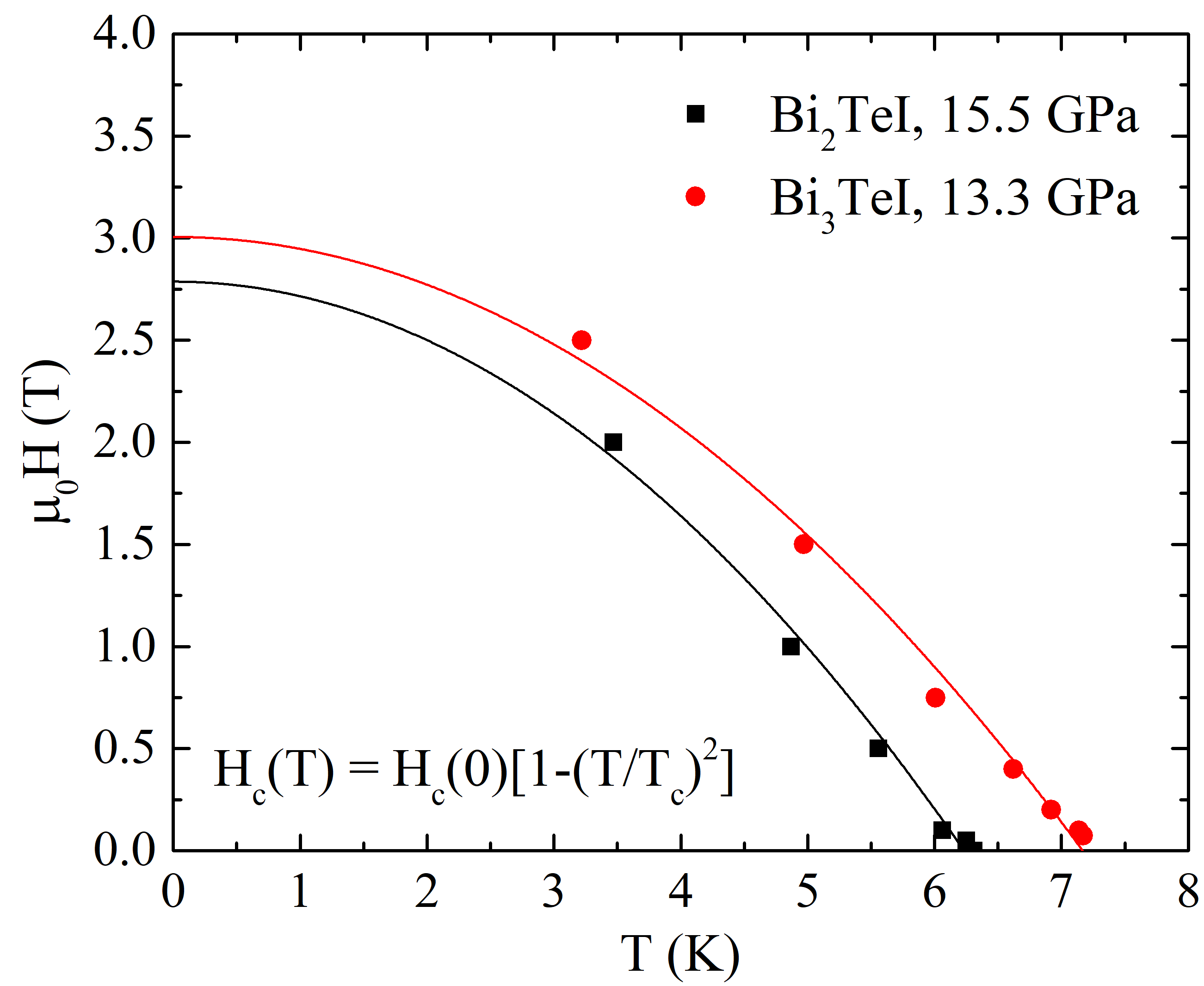}
    \caption{Magnetic field vs temperature for both samples. Data points are taken from the onset of superconductivity measured in $\chi^\prime$.}
 \label{fig:acs_critical_field}
\end{figure}

An external magnetic field was applied to the samples that were studied via ac magnetic susceptibility in order to investigate the critical fields of Bi$_2$TeI and Bi$_3$TeI (Fig.~\ref{fig:acs_critical_field}).
Empirical parabolic fits yield H$_c$(0) = \SI{2.8 \pm 0.1}{\tesla} for Bi$_2$TeI at \SI{15.5}{\giga \pascal}, and H$_c$(0) = \SI{3.0 \pm 0.1}{\tesla} for Bi$_3$TeI at \SI{13.3}{\giga \pascal}. 
A Werthamer-Helfand-Hohenberg analysis $H_{c2}(0) = -0.7T_c(dH_{c2}/dT)\rvert_{T=T_c}$ yields a zero temperature critical field of \SI{3.1 \pm 0.03}{\tesla} for Bi$_2$TeI at \SI{15.5}{\giga \pascal}, and \SI{3.16 \pm 0.03}{\tesla} for Bi$_3$TeI at \SI{13.3}{\giga \pascal}~\cite{werthamer_temperature_1966}.
The critical fields for these compounds are well below the weak coupling BCS paramagnetic limit, $\mu_0H_p^{BCS} = 1.84T_c$, which yields \SI{11.4}{\tesla} for Bi$_2$TeI at \SI{15.5}{\giga \pascal}, and \SI{12.9}{\tesla} for Bi$_3$TeI at \SI{13.3}{\giga \pascal}.

\section{X-ray diffraction}
\begin{figure}
    \includegraphics[width=\columnwidth]{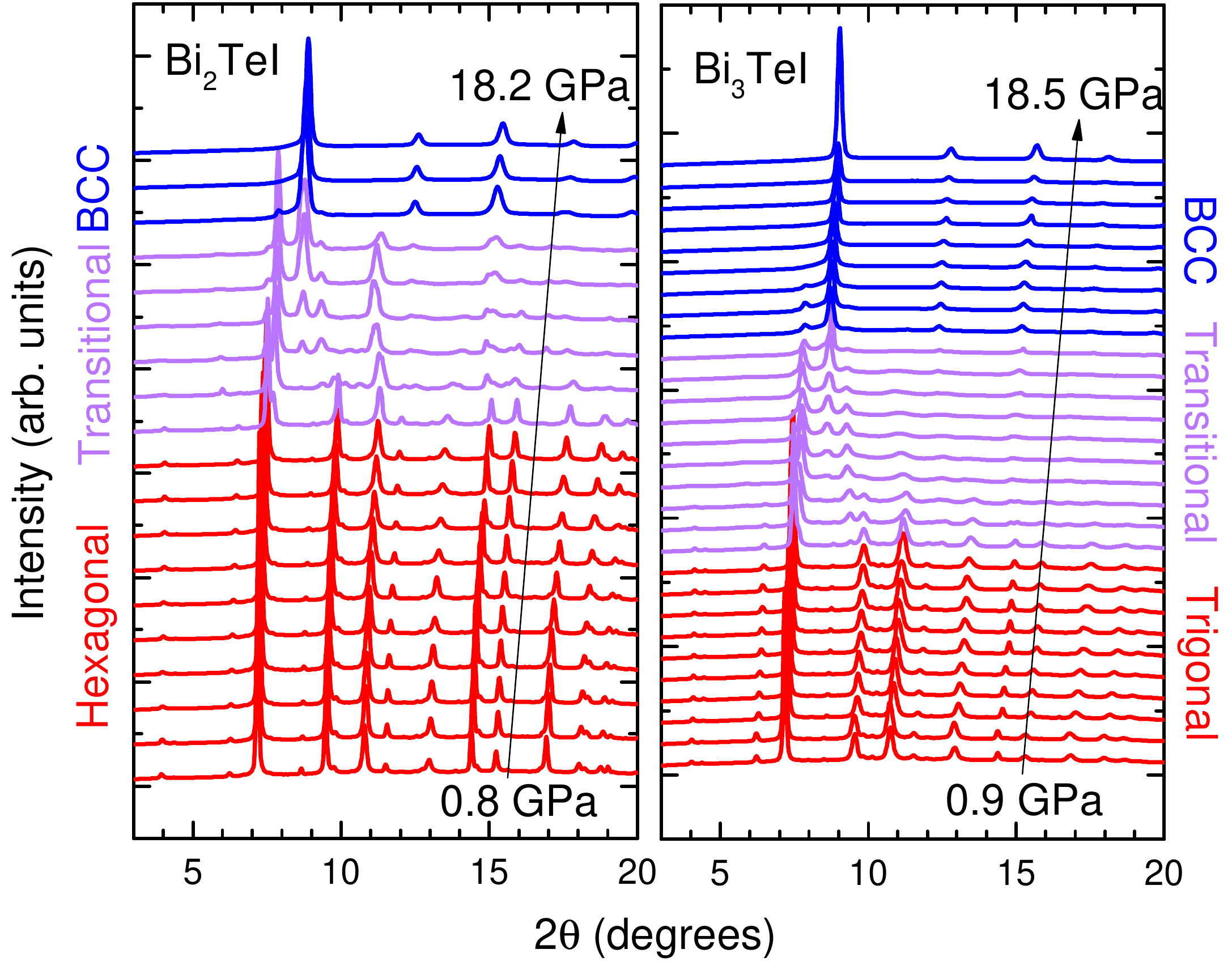}
    \caption{X-ray diffraction patterns for Bi$_2$TeI and Bi$_3$TeI.
    Data are offset vertically for clarity.
    Both compounds undergo structural transitions as pressures increase.
    Bi$_2$TeI exists in a hexagonal phase at ambient pressure, while Bi$_3$TeI has a trigonal structure.
    However a transitional phase emerges starting at about \SI{7}{\giga \pascal} in Bi$_2$TeI and \SI{5.5}{\giga \pascal} in Bi$_3$TeI.
    The structure fully transforms to a $bcc$ phase above \SI{13}{\giga \pascal} in Bi$_2$TeI and \SI{10}{\giga \pascal} in Bi$_3$TeI.}
 \label{fig:Bi2TeI_Bi3TeI_XRD}
\end{figure}

Figure~\ref{fig:Bi2TeI_Bi3TeI_XRD} summarizes the results of the high pressure x-ray diffraction measurements for both compounds.
As pressure increases from zero to \SI{18}{\giga \pascal}, Bi$_2$TeI passes through three distinct phases.
At low pressure, it adopts a hexagonal structure (space group C12/m1), which transforms into an unidentified transitional phase at approximately \SI{7}{\giga \pascal}.
Above \SI{13}{\giga \pascal}, a third phase appears, and this phase has a remarkably simple diffraction pattern.
Bi$_3$TeI behaves similarly under pressure.
At low pressure, it possesses a trigonal structure (space group R3m), and the structure shifts into a transitional phase at around \SI{5.5}{\giga \pascal}.
Upon increasing pressure above about \SI{10}{\giga \pascal}, Bi$_3$TeI also transitions to a new phase with an extremely simple diffraction pattern.

Attempts to index the high-pressure diffraction patterns of \BBTI{} and \BBBTI{} yield unit cells that are too small to contain even a single formula unit.
Small unit cells can occur for complex formula units if each lattice site hosts multiple types of atom - i.e. if the structure is disordered.
We find that the high pressure phases of both \BBTI{} and \BBBTI{} are well explained by a simple $bcc$ lattice where the the Bi, Te, I atoms are randomly distributed.
The appearance of a disordered $bcc$ structure in both materials at high pressure is not entirely unexpected, given that the same kind of structure is known to appear in a variety of Bi-Te compounds~\cite{loa_atomic_2016,stillwell_superconducting_2016}.
When pressure was unloaded, this $bcc$ phase persisted until about \SI{6}{\giga \pascal}.
\begin{figure}
    \includegraphics[width=\columnwidth]{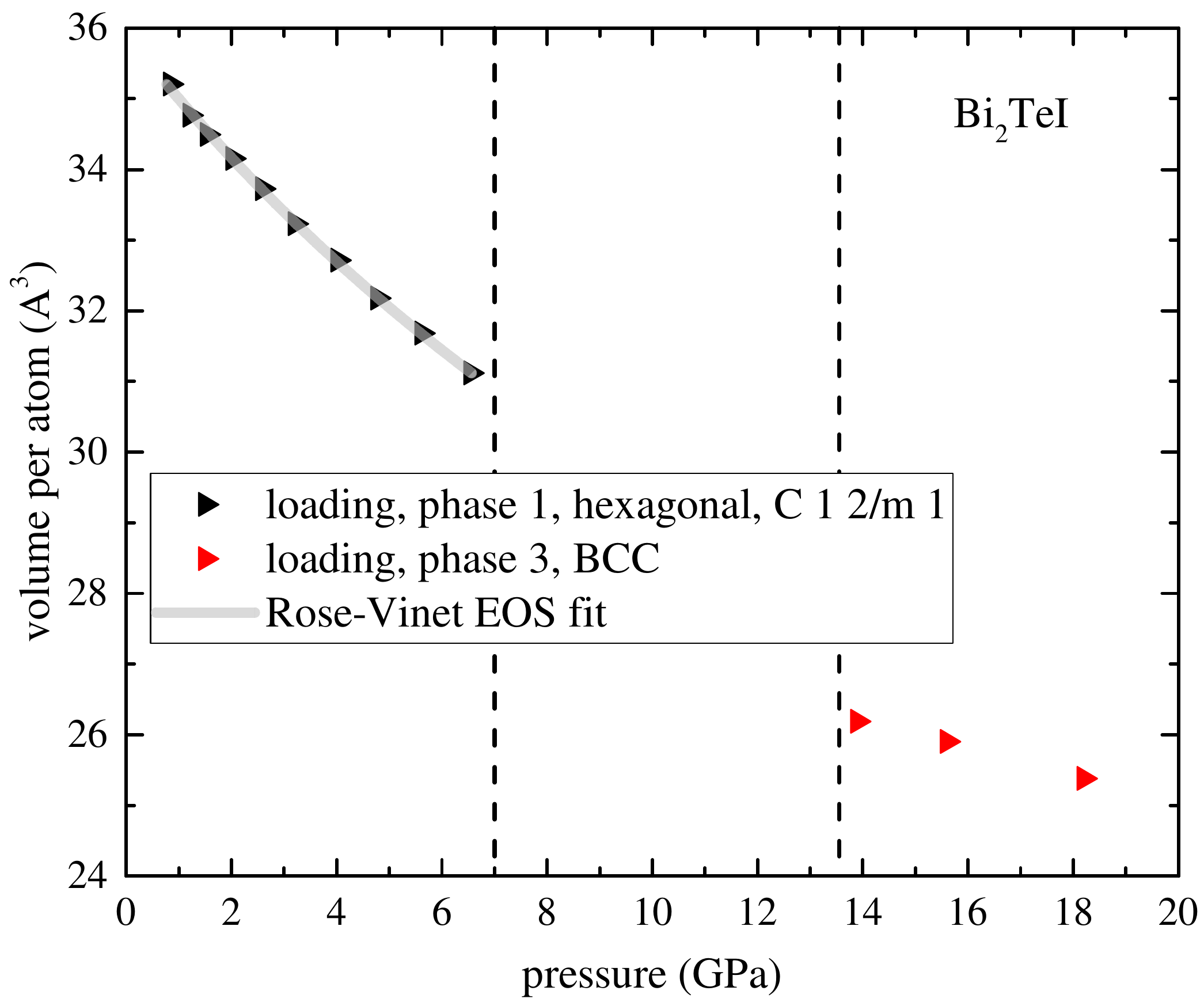}
    \caption{Volume per atom versus pressure for Bi$_2$TeI.  A second, transitional phase appears above about \SI{7}{\giga \pascal}, and a third, BCC phase appears above about \SI{13.5}{\giga \pascal}.  The grey line is a Rose-Vinet Equation of State fit to the low pressure phase data, which results in the parameters B$_0$ = \SI{37 \pm 1.8}{\giga \pascal}, B$_0^\prime$ = \SI{2.99 \pm 0.56}{}, and V$_0$ = \SI{574.73 \pm 0.95}{\cubic \angstrom}}
 \label{fig:Bi2TeI_VvsP}
\end{figure}

\begin{figure}
    \includegraphics[width=\columnwidth]{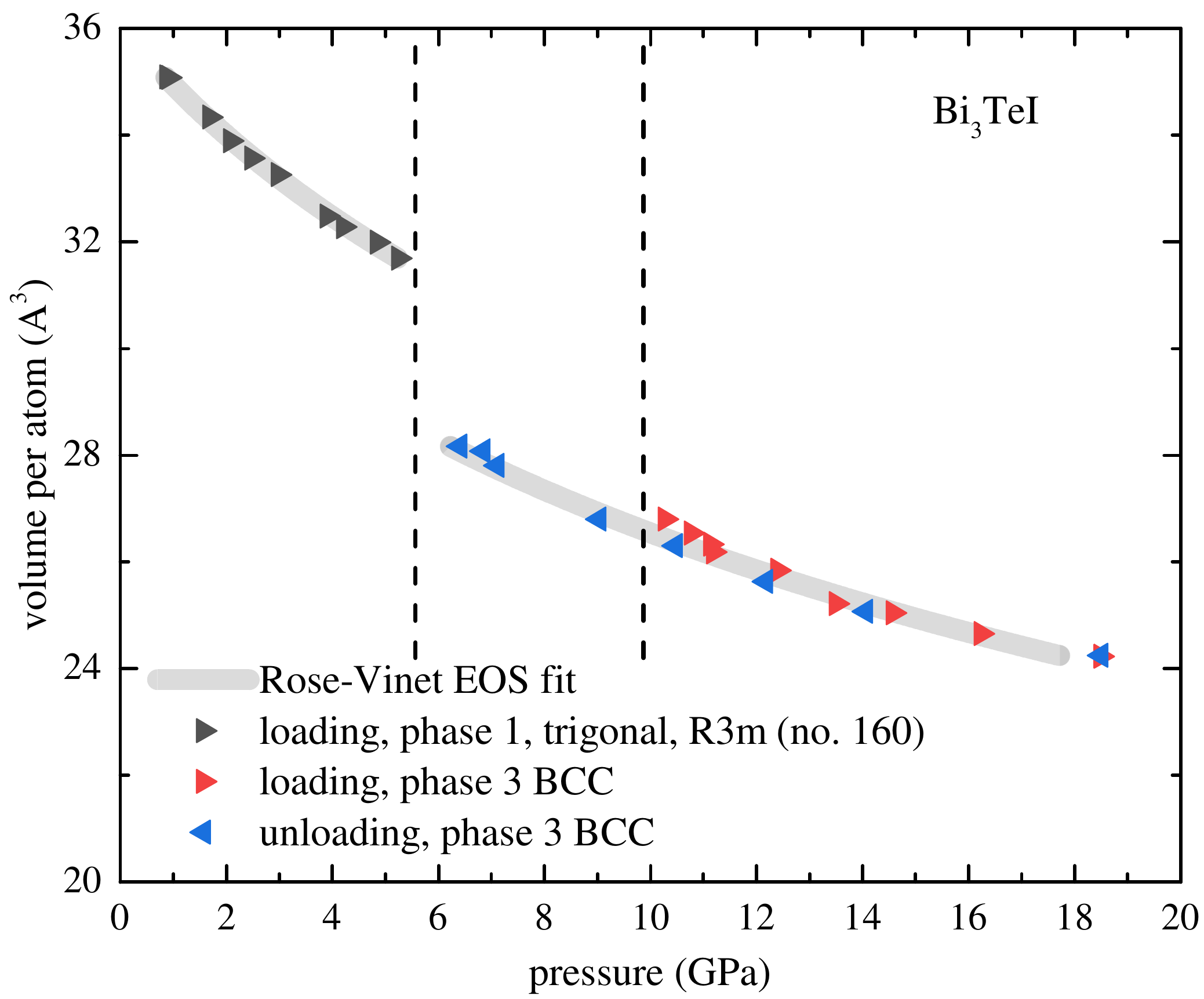}
    \caption{Volume per atom versus pressure for Bi$_3$TeI.  Vertical dashed lines indicate phase transitions. When applying pressure, a transitional phase appears (phase 2) above about \SI{5.5}{\giga \pascal}, which becomes a BCC phase (phase 3) above about \SI{10}{\giga \pascal}.  On unloading, phase 3 persists until about \SI{6}{\giga \pascal}.}
 \label{fig:Bi3TeI_VvsP}
\end{figure}

The structural changes are also illustrated in Figs.~\ref{fig:Bi2TeI_VvsP} and \ref{fig:Bi3TeI_VvsP}, which show the pressure dependent volume per atom of Bi$_2$TeI and Bi$_3$TeI, respectively.
The grey line in Fig.~\ref{fig:Bi2TeI_VvsP} is a Rose-Vinet Equation of State (EOS) fit to the low pressure data, which gives the parameters B$_0$ = \SI{37 \pm 1.8}{\giga \pascal}, B$_0^\prime$ = \SI{2.99 \pm 0.56}{}, and V$_0$ = \SI{574.73 \pm 0.95}{\cubic \angstrom}.
The high pressure EOS was not fitted due to the small number of data points.

Two Rose-Vinet Equation of State fits were performed on the Bi$_3$TeI data, one for each of the high and low pressure phases, and are indicated in Fig.~\ref{fig:Bi3TeI_VvsP} by light grey lines.
In the fit to the low pressure data, V$_0$ was fixed to \SI{540.7}{\cubic \angstrom} based on the known value of the ambient pressure cell volume as reported in~\cite{zeugner_modular_2017}.
This fit gives the values B$_0$ = \SI{29.5 \pm 1}{\giga \pascal} and B$_0^\prime$ = \SI{5 \pm 0.6}{}.
At high pressure, there were not enough data points to obtain reasonable values from the Rose-Vinet fit on its own.
In order to make a rough estimate of B$_0$ and V$_0$, B$_0^\prime$ was fixed to 4.78, which is the value of B$_0^\prime$ for the high pressure phase of Bi$_2$Te, taken from~\cite{stillwell_superconducting_2016}.
The result of this fit gives B$_0$ = \SI{30 \pm 3}{\giga \pascal}.

\begin{figure}
    \includegraphics[width=0.95\columnwidth]{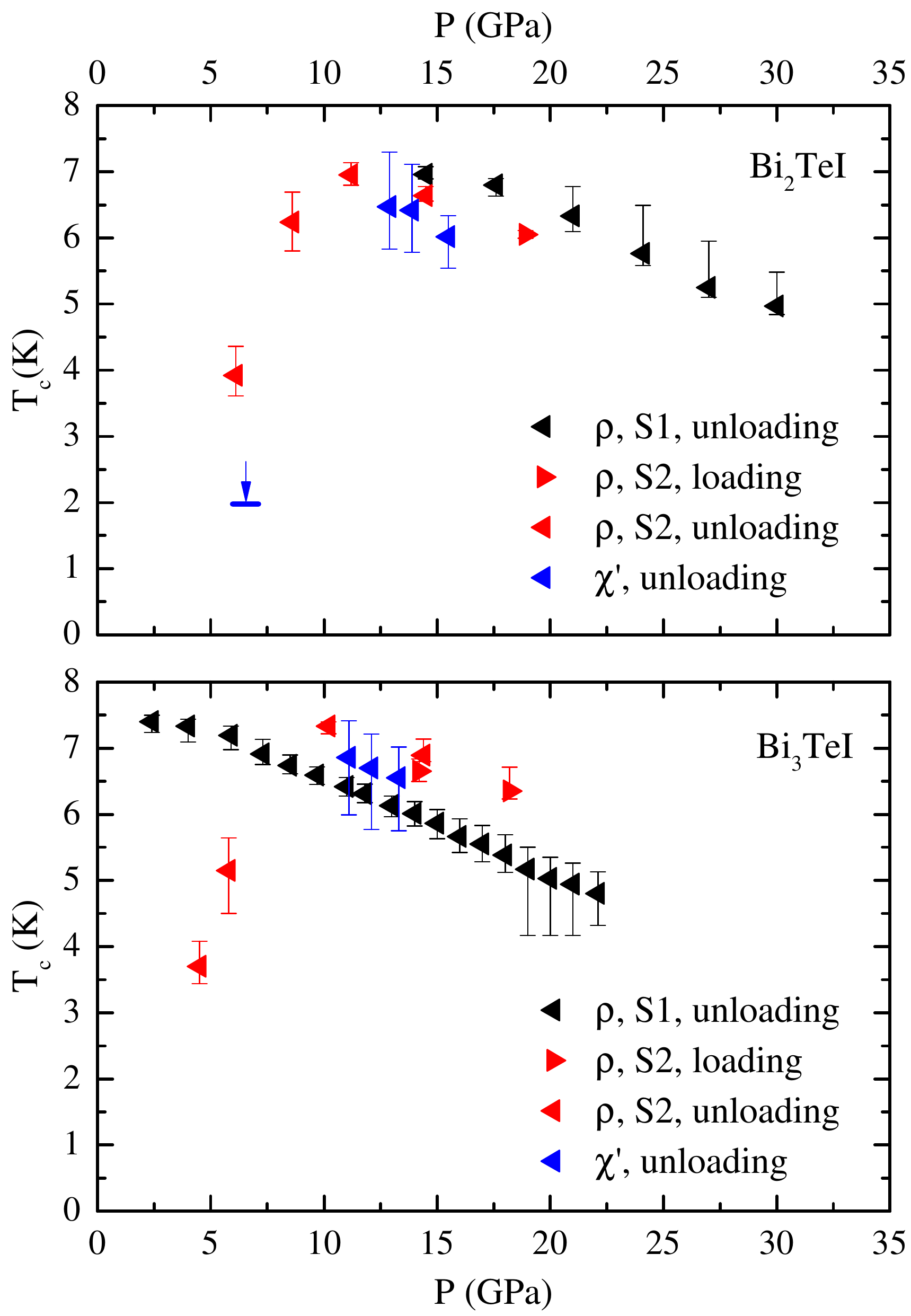}
    \caption{Superconducting phase diagrams for \BBTI{} (top) and \BBBTI{} (bottom) for all samples measured at low temperature. The horizontal line with an arrow indicates the lack of a superconducting signal in our ac susceptibility measurements.}
 \label{fig:phase_diagram}
\end{figure}

Phase diagrams for both compounds are presented in Fig.~\ref{fig:phase_diagram}.
The results show that the high pressure $bcc$ phase exhibits a negative $dT_c/dP$.
Upon pressure unloading, superconductivity persists, suggesting that the $bcc$ structure is metastable.

\section{Discussion}
\begin{figure}
    \includegraphics[width=\columnwidth]{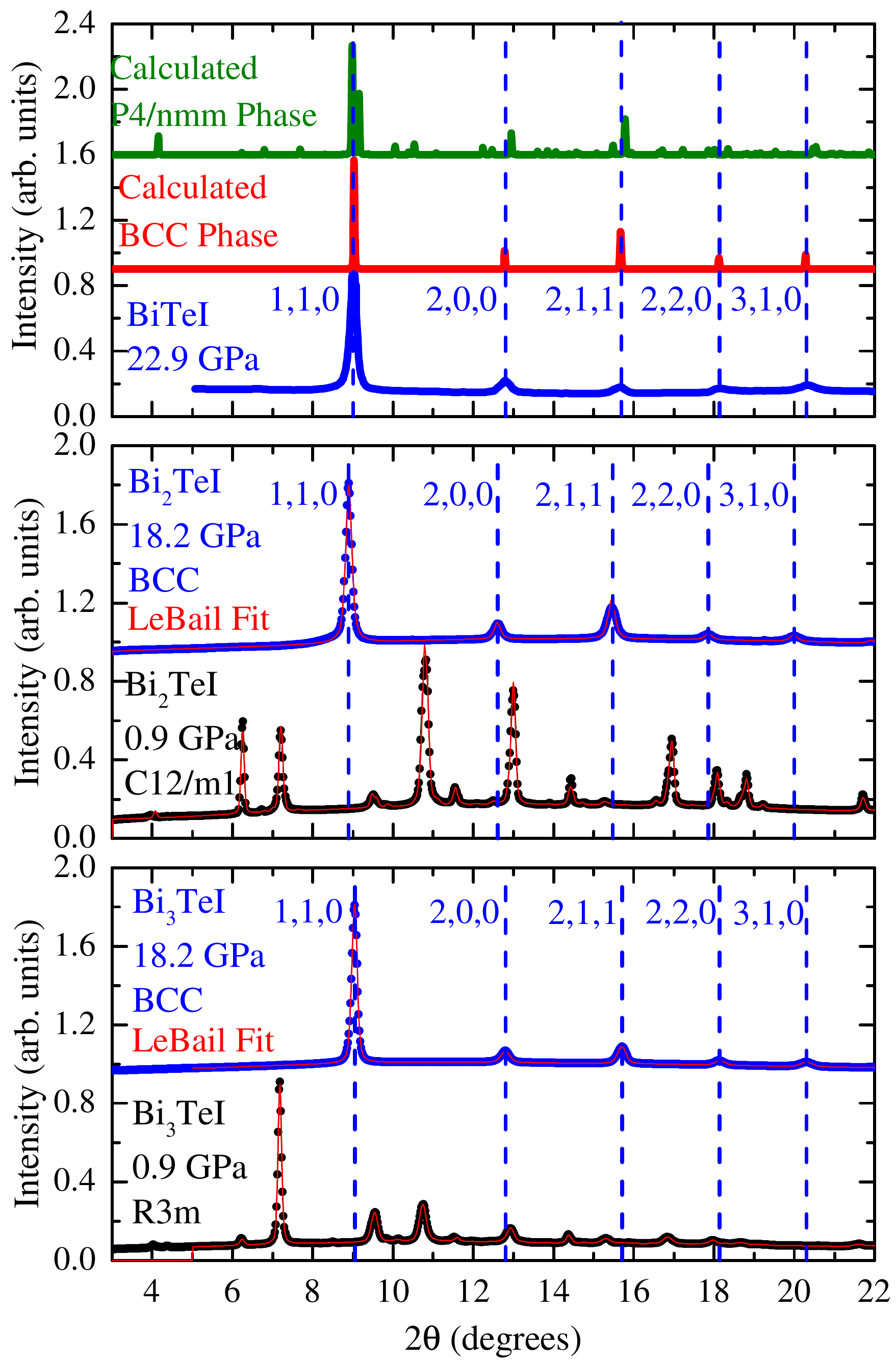}
    \caption{(a) X-ray diffraction pattern for BiTeI, taken from the work of Chen et al.~\cite{chen_high-pressure_2013}, shown in blue.  Vertical dashed lines indicate a peak positions predicted by an Im3m structure assuming an x-ray wavelength of \SI{0.4066}{\angstrom} and unit cell of side length a = \SI{3.63232}{\angstrom}.  In red and green, respectively, calculated BCC and P4/nmm phases. (b) \& (c) XRD patterns for Bi$_2$TeI and Bi$_3$TeI, respectively, comparing the high and low pressure phases.}
 \label{fig:BiTeI_XRD}
\end{figure}

Since \BBTI{} and \BBBTI{} are composed of BiTeI layers with Bi-bilayers, we find it useful to compare our high-pressure results to those observed in BiTeI.
In BiTeI, structural transitions occur near \SI{9}{\giga \pascal} and \SI{19}{\giga \pascal} at room temperature~\cite{chen_high-pressure_2013}. 
BiTeI begins to superconduct either in the high-pressure phase II or III~\cite{vangennep_pressure-induced_2017,jin_superconductivity_2017,qi_topological_2017}.
Though Chen et al.~\cite{chen_high-pressure_2013} report that phase III of BiTeI exists in a P4/nmm structure, we find that a $bcc$ (Im3m) structure produces much better agreement with the observed high pressure diffraction peaks, as demonstrated in Fig.~\ref{fig:BiTeI_XRD}(a).
This would also align with other works that suggest similar $bcc$ phases in other bismuth telluride compounds~\cite{loa_atomic_2016}.
Our own x-ray diffraction measurements, shown in Fig.~\ref{fig:BiTeI_XRD}(b) and (c), demonstrate that Bi$_2$TeI and Bi$_3$TeI also assume this same structure in their high-pressure phases.
In BiTeI, T$_c$ reaches a maximum of \SI{5.8}{\kelvin} near \SI{25}{\giga \pascal}~\cite{vangennep_pressure-induced_2017} under decompression at low temperature. 
\BBTI{} achieves a maximum T$_c$ of $\sim$\SI{7}{\kelvin} near \SIrange{11}{14}{\giga \pascal} upon decompression, and \BBBTI{} achieves a maximum T$_c$ of \SI{7.4}{\kelvin} near 2 and \SI{10}{\giga \pascal}, also upon decompression.
The introduction of Bi layers into the BiTeI system tends to increase T$_c$, while H$_c$ for these compounds are all roughly the same (\SIrange{2}{3}{\tesla})~\cite{vangennep_pressure-induced_2017,jin_superconductivity_2017,qi_topological_2017}.
Our studies on \BBTI{} and \BBBTI{} show systematic differences in T$_c$ which are similar to those seen in the Bi$_2$Te, Bi$_2$Te$_3$ and Bi$_4$Te$_3$~\cite{jeffries_distinct_2011,stillwell_superconducting_2016}, where the addition of Bi-bilayers increases T$_c$, and maximum values of T$_c$ tend towards the maximum T$_c$ value observed in elemental Bi ($\sim$\SI{8}{\kelvin})~\cite{lotter_evidence_1988,valladares_possible_2018}.
\begin{figure}
    \includegraphics[width=\columnwidth]{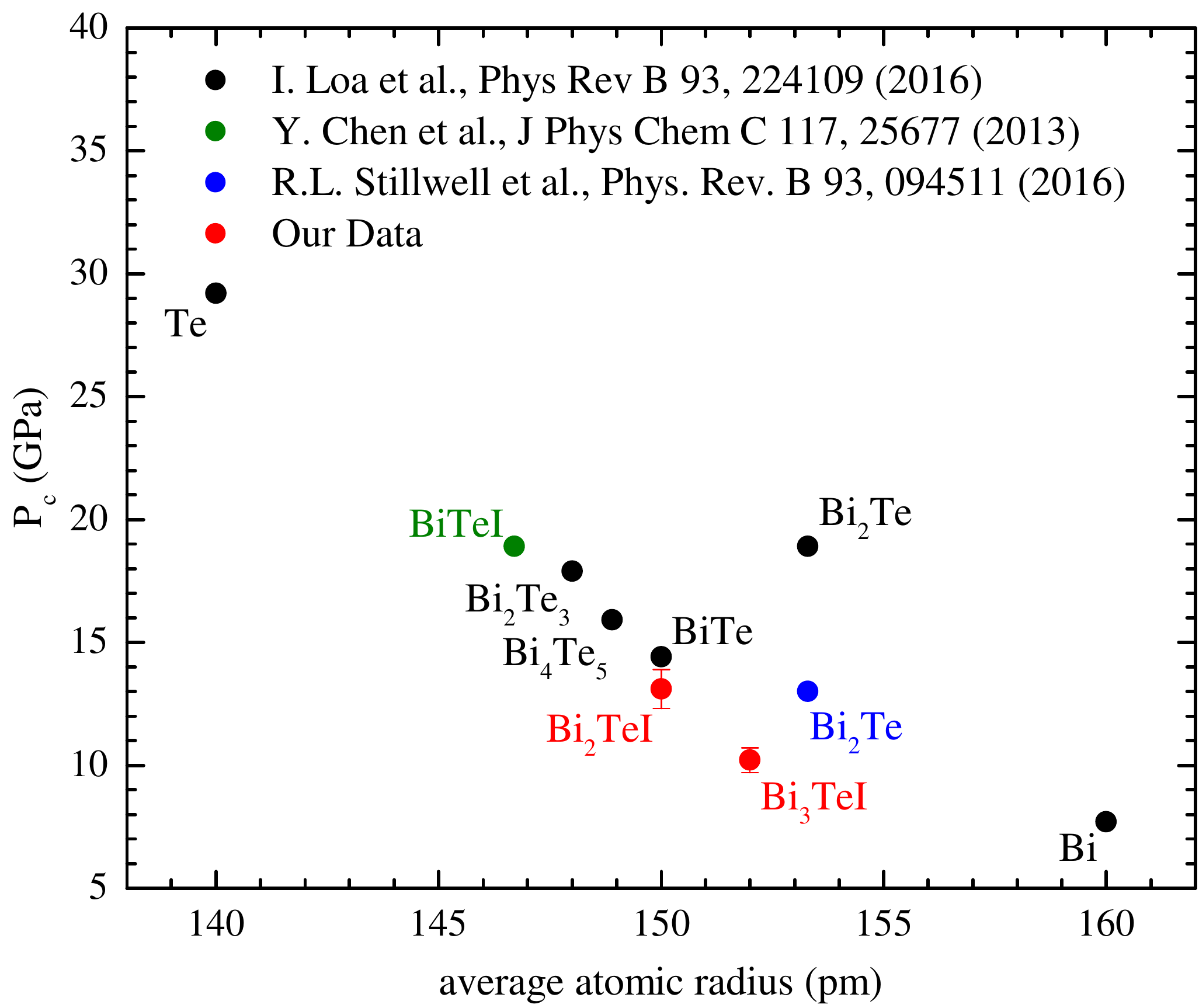}
    \caption{The relation between average atomic radius and critical pressure required to induce the structural transition to disordered $bcc$ for Bi$_2$TeI and Bi$_3$TeI compared to data gathered by I. Loa et al.~\cite{loa_atomic_2016}, Y. Chen et al.~\cite{chen_high-pressure_2013}, and R. L. Stillwell et al.~\cite{stillwell_superconducting_2016}  Average atomic radii were taken from J.C. Slater.~\cite{slater_atomic_1964}}
 \label{fig:PcvsR}
\end{figure}

A comparison between the critical pressure for the transition to the disordered $bcc$ structure and the average atomic radius for several bismuth telluride compounds is presented in Fig.~\ref{fig:PcvsR}.
Volumes were calculated from covalent atomic radii provided by J.~C.\ Slater~\cite{slater_atomic_1964}.
Figure~\ref{fig:PcvsR} includes data from I.\ Loa et al.~\cite{loa_atomic_2016}, and also includes BiTeI data from Y. Chen et al.~\cite{chen_high-pressure_2013} (based on our $bcc$ fit of their diffraction data), Bi$_2$Te data from R. L. Stillwell et al.~\cite{stillwell_superconducting_2016}, and our own data on the compounds Bi$_2$TeI and Bi$_3$TeI.
A clear trend is observed where the critical pressure for the transition to the disordered $bcc$ structure drops with increasing average atomic radius.

\section{Conclusion}

We report electrical resistivity and
magnetic susceptibility measurements on \BBTI{} and \BBBTI{} to pressures
as high as $\sim$\SI{30}{\giga \pascal}. 
Both samples exhibit pressure-induced structural transitions indicated by sharp drops in the resistivities as a function of pressure.
X-ray diffraction measurements reveal a disordered $bcc$ structure emerges at high pressure in both compounds.
Magnetic susceptibility measurements show features consistent with 100\% superconducting shielding, and thus confirm the bulk nature of the observed superconductivity.
The results indicate that Bi-Te and Bi-Te-I compounds, while structurally distinct at ambient pressure, eventually transform into a universal superconducting $bcc$ alloy phase at high pressure.
Bi-Te-I compounds effectively become multi-principal element alloys (closely related HEA ``high entropy alloys'') at high pressure and join a large family of such HEA-type superconducting materials~\cite{Mizuguchi2021}.

\section*{Acknowledgments}
Initial experiments supported by National Science Foundation (NSF) CAREER award DMR-1453752.
Data analysis supported by NSF DMREF-2118718.
High pressure technique development was partially supported by a National High Magnetic Field Laboratory User Collaboration Grant.
The National High Magnetic Field Laboratory is supported by the NSF via Cooperative agreement No.\ DMR-1157490, the State of Florida, and the U.S.\ Department of Energy.
RB was supported by NSF DMR-1904361.
We thank Yue Meng for experimental assistance.
Portions of this work were performed at HPCAT (Sector 16), Advanced Photon Source (APS), Argonne National Laboratory.
HPCAT operations are supported by DOE-NNSA's Office of Experimental Sciences.
The Advanced Photon Source is a U.S.\ Department of Energy (DOE) Office of Science User Facility operated for the DOE Office of Science by Argonne National Laboratory under Contract No.\ DE-AC02-06CH11357.

\bibliography{Bi-Te-I}
\end{document}